
\input phyzzx
\pubnum{05}
\pubtype={}
\titlepage
\title{ Uniqueness of the Axionic Kerr Black Hole}
\author{Rue-Ron Hsu, Green Huang and Wei-Fu Lin}
\address{Department of Physics, National Cheng Kung University,}
\address{Tainan, Taiwan, 70101, Republic of China}
\vskip 1.5 in
\hfill {PACS:04.20.Jb,04.50.+h}
\abstract
Under the axisymmetry and under the invarance of simultaneous inversion of
time and azimuthal angle, we show that
the axionic Kerr black hole is the ${\it unique}$ stationary
solution of the minimal coupling theory of gravity and the Kalb-Ramond
field, which has a regular event horizon, is asymptotically flat and
has a finite axion field strength at event horizon.
\endpage

\chapter{Introduction}
Recently, a lot of interest has been aroused in the theory in which
gravity is coupled to the Kalb-Ramond field~\refmark{1-5}. The only
static, spherically symmetric solution to the minimal coupling theory,
which has a regular event horizon, is asymptotically flat and has a
finite axionic charge $\it q$, is the Schwarzschild
solution with vanishing axion field strength but nonvanishing
Kalb-Ramond field.\refmark{1,4}
{}~Since the axion field stength vanishes
everywhere and the axion charge can be detected not by a point
particle but by a string in a process analogous
to the Aharonov-Bohm effect, the axionic charge is regarded as a
quantum hair of black hole\refmark{6}. Studying the axisymmetric field
equation of the theory, we found an axionic Kerr black hole solution with
non-vanishing axion field strength for
the case of slow rotation~$(a^2 < M^2)$
{}~\refmark{5}. In addition to mass and angular momentum, the solution
indicates that there exists a new classical hair called axionic hair,
which can be detected by a test particle directly without resorting to
the stringy Aharonov-Bohm process.

In this paper, we show that,
under the axial symmetry and under the invarance
of simultaneous inversion of time and azimuthal angle, the axionic
Kerr black hole is the ${\it unique}$ stationary  solution of the
minimal coupling theory of garvity and Kalb-Ramond field.
This black hole has a regular event horizon, is asymptotically flat and
has a finite axion field strength at event horizon.

\chapter{The stationary axisymmetric Einstein-Kalb-Ramond equation}
The minimal coupling theory of Kalb-Ramond field and gravity is described
by the action
$$\eqalign{ I\, = \, & \int d^4x \, \sqrt{-g} \, \bigl( \,
	    {1\over{2\kappa}}
	    R \, - \, {1\over 6} H_{\mu\nu\lambda}
	    \,H^{\mu\nu\lambda} \, \bigr), \cr} \eqno(1) $$
in which
$$\eqalign{ H_{\mu\nu\lambda} \, = & \, \partial_\mu B_{\nu\lambda} \, +
	    \,	 \partial_\nu \, B_{\lambda\mu} \, +
	    \, \partial_\lambda \, B_{\mu\nu}\, \cr}\eqno(2) $$
is the field strength of
Kalb-Ramond field $B_{\mu\nu}$, and the field equations are
the Einstein-Kalb-Ramond (EKR) equation :
\foot{We shall use the conventions given in reference[7].}
$$\eqalign{ R_{\mu\nu} \, - \, {1\over 2} {g_{\mu\nu}}R \, = & \,- \kappa
	    \,	\bigl( \,H_{\mu\lambda\rho} H_\nu\,^{\lambda\rho}\,-\,
	    {1\over 6} {g_{\mu\nu}} H_{\lambda\rho\sigma}
	    H^{\lambda\rho\sigma} \, \bigr) \cr} \eqno(3) $$
$$\eqalign{ \partial_\sigma \, \bigl( \, \sqrt{-g} H^{\sigma\mu\nu}
	    \, \bigr)	\, = & \,0\,. \cr} \eqno(4) $$

To describe a stationary axisymmetric spacetime, it is
convenient to take the time $t$ and the azimuthal angle $\phi$ about
the axis of symmetry as two of the coordinates.  The stationary
and the axisymmetric character of the spacetime require that the
metric be independent of $t$ and $\phi$ \refmark{7,8}.
Besides stationarity
and axisymmetry, it is also required that the spacetime is invariant
under the simultaneous inversion of the time $t$ and the angle $\phi$,
{\it i.e.}, to the transformation $t \to -t$ and $\phi \to -\phi$, for a
purely rotation.  Hence, under the assumption made, the metric \refmark{7}
components
$$g_{tr}=g_{t\theta}=g_{\phi r}=g_{\phi\theta}=0,$$
where $r$ and $\theta$ are the two remaining spatial coordinates.
Moreover, a further reduction in the form of the metric can be
achieved by using the theorem --- a two dimensional space,
with a positive or a negative definite signature, can always
be brought to diagonal form by a coordinate transformation.
Therefore, the stationary axisymmetric metric is
$$\eqalign{ ds^2  =& e^{2A(r, \theta)} dt^2  -	e^{2B(r,\theta)}
    \bigl(  d \phi - \omega {(r, \theta)} dt \bigr) ^2 - e^{2C(r, \theta)}
     dr^2 - e^{2D(r, \theta)} d \theta ^2 . \cr} \eqno(5) $$
Here, we introduce the tetrad frame with basis 1-forms
$$\omega^0=e^A dt,\omega^1=e^B(d\phi-\omega dt),\omega^2=e^C dr,
 \omega^3=e^D d\phi.$$

We also impose the stationarity, axismmetry and simultaneous inversion
invariance upon the Kalb-Ramond field.	Analogous to the metric
components, the corresponding non-vanishing
tetrad components of~$B_{\mu\nu}$~ and~$H_{\mu\nu\lambda}$
are \refmark{8}
$$\eqalign{B_{01}\, =& \,-\, B_{10}\, \equiv \,E(r, \theta) \cr
	   B_{23}\, =& \,-\, B_{32}\, \equiv \,F(r, \theta) \,~,\cr}
	   \eqno(6) $$
$$\eqalign{H_{012}\,=&\,e^{-A-B-C} (E(r,\theta) e^{A+B}),_r \cr
	   H_{013}\,=&\,e^{-A-B-D} (E(r,\theta) e^{A+B}),_\theta ~.\cr
	   } \eqno(7) $$
The component $B_{23}$ can be regarded as a gauge freedom that does not
contribute the field strength $H_{\mu\nu\lambda}$.

Thus, the stationary axisymmetric EKR equations are
$$\eqalign{{1\over2} e^{2B-2A}\, \lbrack \, e^{-2C}(\omega,_r)^2\, +
	   \, e^{-2D} (\omega,_\theta)^2\, \rbrack\,\cr =\, e^{-2C}\,
	   \lbrack \, A, _r, _r\,+ \, A, _r \,
	   (A+B-C+D),_r\,\rbrack\,\cr+\, e^{-2D}\,\lbrack\,
	   A,_\theta,_\theta\,+\, A,_\theta
	   \,(A+B+C-D),_\theta\,\rbrack \cr} \eqno(8) $$
$$\eqalign{\bigl(\, e^{3B-A-C+D}\,\omega,_r\,\bigr),_r\, + \, \bigl(\, e^{
   3B-A+C-D}\,\omega,_\theta\,\bigr),_\theta\, = & \,0 \cr} \eqno(9) $$
$$\eqalign{{-1\over2} e^{2B-2A}\, \lbrack\,
	  e^{-2C}(\omega,_r)^2\, +\, e^{-2D}
	  (\omega,_\theta)^2\, \rbrack\, \cr =\, e^{-2C}\,
	  \lbrack\, B,_r,_r\,+\, B,_r
	  \, (B+A-C+D),_r\, \rbrack\, \cr +e^{-2D}\, \lbrack\,
	  B,_\theta,_\theta\,+\,
	  B,_\theta\,(B+A+C-D),_\theta\, \rbrack } \eqno(10) $$
$$\eqalign{{-1\over2} \omega,_r\omega,_\theta e^{2B-2A}\,
	  +\, (A+B),_r,_\theta
	  \,-\, (A+B),_r C,_\theta\, \cr -\, (A+B),_\theta
	  D,_r\, +\, A,_rA,_\theta\,
	  +\, B,_rB,_\theta\, \cr =\, 2\kappa
	  e^{-2A-2B} (E e^{A+B}),_r (E e^{A+B}),_
	  \theta \cr} \eqno(11) $$
$$\eqalign{e^{-2C}\,\lbrack&\, A,_r(B+D),_r\,+\, B,_rD,_r\, \rbrack\, +\,
	  {1\over4} e^{2B-2A}\, \lbrack\, e^{-2C}(\omega,_r)^2\,-\,
	  e^{-2D}(\omega ,_\theta)^2\, \rbrack \cr
	  +&\, e^{-2D} \, \lbrack\, (A+B),_\theta,_\theta\,
	  +\, (B+A),_\theta(A-D),_\theta\, +\, (B,_\theta)^2\, \rbrack \cr
	  =&\, -\kappa
	  \, \lbrack\, -e^{-2A-2B-2C}((Ee^{A+B}),_r)^2\,
	  +\,e^{-2A-2B-2D}((Ee^{A+B}),_
	  \theta)^2\, \rbrack \cr } \eqno(12) $$
$$\eqalign{e^{-2D}\, \lbrack&\, A,_\theta(B+C),_\theta\, +\, B,_\theta
	   C,_\theta \, \rbrack\,-\, {1\over4}e^{2B-2A}\,
	   \lbrack\, e^{-2C}(\omega,_r)^2\,-\,
	   e^{-2D}(\omega,_\theta)^2 \, \rbrack \, \cr
	   +&\, e^{-2C}\, \lbrack\, (A+B),_r
	   ,_r\,+\, (B+A),_r(A-C),_r\,+B,_rB,_r\, \rbrack\, \cr
	   =&\, -\kappa\, \lbrack\, e^{-2A-2B-2C}((Ee^{A+B}),_r)^2
	   \,-\, e^{-2A-2B-2D}((Ee^{A+B}),_\theta)^2\,
	   \rbrack\, \cr } \eqno(13) $$
$$\eqalign{\bigl(\, e^{-A-B-C+D}\,(Ee^{A+B}),_r\, \bigr)\, ,_r\,
	   +\, \bigl(\, e^{-A-B+C-D}\,
	   (Ee^{A+B}),_\theta\, \bigr)\, ,_\theta\, =\, 0\,. \cr}
	   \eqno(14)  $$

\chapter{Uniqueness proof of the axionic Kerr black hole}
It is quite amazing that eqs.(8)-(10) are identical to eqs.(5)-(7)
in the page 274 of reference [7].  Such that,
following the same procedure in reference [7],
we take a choice of gauge freedom
$$\eqalign{e^{2D-2C}=\, \Delta,\cr}\eqno(15) $$
this is consistent with the existence of an event horizon,
and in accordance with the Robinson's theorem~\refmark{7,9},
$$\eqalign{e^{2A}\,=&\, \Delta \rho^2\, \over{\Sigma^2}\cr e^{2B}\,=&\,
   \delta \Sigma^2\, \over {\rho^2}\cr \omega\,=&\,{2aMr\, \over{\Sigma^2}}
   , \cr } \eqno(16) $$
in which
$$\eqalign{ \Delta\,=&\, r^2 \,-\, 2Mr\,+\,a^2 \cr \rho^2\,=&\, r^2 \,+\,
	    a^2 \mu^2 \cr \Sigma^2\,=&\, (r^2\,+\,a^2\,)^2\,-\,
	    a^2 \Delta \delta \cr
	    \mu\,=&\,cos\theta \cr \delta\,=&\, sin^2\theta \,. \cr }
	    \eqno(17) $$
is the {\it unique} solution of eqs.(8)-(10) which satisfy the conditions
of asymptotical flat, regular event horizon and non-singular outside
of the horizon.
The integration constants $M\,$ and$\, a$ parametrize the mass and spin
angular momentum per mass of the rotating black hole.

Now, the remaining problem is to solve eqs.(11)-(14).
Plugging eq.(16) into eq.(14), we end up with
$$\eqalign{{1\over{\sqrt{\delta}}}(\sqrt{\Delta} E),_r,_r\, +\,
	   {1\over{\sqrt{
	   \Delta}}} (\sqrt{\delta} E),_\mu ,_\mu \,=\, 0 .  \cr}
	   \eqno(18) $$
In reference[5], we found an asymptotical-flat solution
$$\eqalign{ E(r,\theta )\,=\, {\alpha \mu \,+\, \beta \over {\sqrt{\Delta
   \delta}}} , \cr} \eqno(19) $$
in which $\alpha$ and $\beta$ are integration constants. Here, we could
find the general solution of~$E(r,\theta)$~ for the slowly rotating case
$(a^2 < M^2)$, and show that eq.(19) is an $\it unique$ solution of eq.(18)
which satisfies the asymptotical-flat condition.

Replacing~$E(r,\theta)$~by
$$\eqalign{E(r,\theta)~\sim~{1\over {\sqrt{\Delta\delta}}}U(r)\Theta(\mu),
	   \cr} \eqno(20)$$
we end up with two separated differential equations
$$\eqalign{\Delta(U(r)),_r,_r-~\ell (\ell + 1)U(r) \,= & \, 0~,\cr}
	   \eqno(21)$$
$$\eqalign{\delta(\Theta(\mu)),_\mu,_\mu +~ \ell (\ell + 1)\Theta(\mu)\,
	   =&\,0~ .\cr}\eqno(22)$$
Using the Frobenius' method, the regular solutions of eq.(22)
are $\Theta_\ell$ polynomials
$$\Theta_\ell = \cases{ \alpha' \mu+ \beta'  ~~~~~~~~~~~~~~~~~~~~~~ for~~
			\ell = 0 \cr
			\sum_{n=0}^m \, a_{2m-1}^{2n} \mu^{2n}~~~~~~~~ for
			~~ \ell = 2m-1~~~~ m \ge 1 \cr
			\sum_{n=0}^m \, a_{2m}^{2n+1} \mu^{2n+1}~~~~~ for~~
			\ell = 2m~~~~~~~~~m \ge 1 \cr} \eqno(23)$$
where
$$\eqalign{&a_{2m-1}^0 = a_{2m}^1 = 1 \cr
	   &a_{2m-1}^{2n} = {1\over(2n)!} \prod_{k=1}^n
	   [(2k-2)(2k-3)-2m(2m-1)]
	      ~~~~~~~ for ~~~ 1 \le n \le m \cr
	   &a_{2m}^{2n+1} = {1\over{(2n+1)!}} \prod_{k=1}^n
	   [(2k-1)(2k-2)-(2m+1)2m]
	     ~~~ for ~~~ 1 \le n \le m \cr} $$
and $\alpha', \beta'$ are integration constants of the the trivial
solution $\Theta_0$.

Here, $\ell$ should be an integer to get convergent series for
all $\mu$ in the regime~$[-1,1]$. Since negative integer values of
$\ell$
would simply give solutions already obtained for positive $\ell$'s,
it is customary to restrict $\ell$ to be non-negative values. For
instance, the first few $\Theta_\ell$ polynomials are
$$\eqalign{\Theta_0= & \alpha' \mu+\beta' , \cr
	   \Theta_1= & 1-\mu ^2, \cr
	   \Theta_2= & \mu-\mu ^3, \cr
	   \Theta_3= & 1-6\mu ^2+5\mu ^4. \cr}$$

Moreover, we rewrite eq.(21) as
$$\eqalign{ \xi (\xi ~+~ \xi_0)U_{, \xi, \xi}~-~\ell (\ell ~+~ 1)U=~0 \cr}
	  \eqno(24)$$
where
$$\eqalign{\xi = & ~ {(r~-~r_+) \over r_+}, \cr
	   \xi_0 = & ~ {(r_+~-~r_-) \over r_+}, \cr}$$
and $r_\pm =~M~ \pm ~\sqrt {M^2~-~a^2}$. The polynomial solutions
$U_\ell $ are
$$\eqalign{U_0(\xi) = & ~~{\alpha''} \xi+ {\beta''} \cr
	   U_{\ell}(\xi)= & ~\sum_{n=0}^{\ell}~b_{\ell}^{n+1}{\xi^{n+1}
		 \over\xi_0^n}\,~~~~~~~~for~~\ell \ge 1 \cr}\eqno(25)$$
in which
$$\eqalign{b_{\ell}^1= & ~1\cr
	   b_{\ell}^{n+1}= & ~{1\over (n+1)!n!} \prod_{k=1}^n [(\ell
		 ~+~1)\ell ~-~ k(k~-~1)]~~~~~for~~1\le n \le \ell~, \cr}$$
and $\alpha'', \beta''$ are integration constants.
The first few $U_{\ell}$ polynomials are
$$\eqalign{U_0=~&\alpha'' \xi+ \beta''  \cr
	   U_1=~&\xi+{\xi^2 \over \xi_0} \cr
	   U_2=~&\xi+{3\over \xi_0}\xi^2+{2\over \xi_0^2}\xi^3 \cr
	   U_3=~&\xi+{6\over\xi_0}\xi^2+{10\over\xi_0^2}\xi^3+
		     {5\over\xi_0^3}\xi^4. \cr}$$
Therefore, the general solution is
$$\eqalign{E(r,\mu)=\sum_\ell {A_\ell \over \sqrt{\Delta\delta}}U_\ell (r)
	       \Theta_\ell(\mu) \cr}\eqno(26)$$
where $A_\ell$ are constants.

Since ${1\over\sqrt\Delta}\sim	r^{-1}$ and
$U_{\ell}(r)\sim \xi^{\ell+1}\sim r^{\ell +1}$ as $r\gg 1$,
the trivial solution eq.(19) is an ${\it unique}$ asymptotically-flat
solution.

After using the replacement
$$\eqalign{e^{C+D}={\rho^2\over{\sqrt\Delta}} f(r,\theta),\cr}\eqno(27)$$
eqs.(16),(19) and the gauge condition eq.(15), those  eqs.(11)-(13)
are reduced to two first order coupled differential equations
$$\eqalign{-{(r-M)\over \Delta} (\ln f),_\mu
		      +{\mu\over\delta}(\ln f),_r =\,0,\cr}\eqno(28)$$
$$\eqalign{(r-M)(\ln f),_r +\mu(\ln f),_\mu
	    = \,-{{2\kappa\alpha^2}\over\Delta}.\cr}\eqno(29)$$
When the asymptotically flat condition
$f(r \to \infty,\theta) \to 1$
is imposed, the solution
$$\eqalign{ f(r,\theta)\,=\, \lbrack\, \mu^2 \,+\, \delta
	    {(r-M)^2 \over{\Delta}}\,
	    \rbrack \,^{-\kappa \alpha^2 \over{a^2-M^2}} \,\cr}
	    \eqno(30) $$
is uniquely determined by straightforward calculation.
Moreover, eqs.(15) and (27) give the metric components
$$\eqalign{e^{2C}=&{\rho^2\over \Delta}f(r,\theta),\cr
	   e^{2D}=&\rho^2 f(r,\theta).\cr}\eqno(31)$$

Base on Robinson's theorem and under the symmetry of
stimultaneous inversion of time and azimuthal angle,
the $\it unique$ axisymmetric black hole is
$$\eqalign{ ds^2\,=\, {\Delta \rho^2\over{\Sigma^2}}\,dt^2\,-\,
	    {\delta\over{\rho^2}}
	    \Sigma^2\,(d\phi\,-\,\omega dt\,)^2\,-\,
	    {\rho^2\over{\Delta}} f
	    dr^2\,-\,\rho^2 f d\theta ^2 , \cr } \eqno(32) $$
and the non-vanishing tetrad compontents
$$\eqalign{ B_{01}\,=\, {\alpha \mu \,+\, \beta \over {\sqrt{\Delta
   \delta}}} , \cr} \eqno(33) $$
where $\delta$ , $\rho^2$ , $\Sigma^2$ , $\delta$ , $\omega , f$,
have been defined in eqs. (17) and (30).
The non-vanishing tetrad components of axionic field strength
$$\eqalign{ H_{013}\,=\, -{\alpha\over{\sqrt{\Delta f} \rho}}\,  \cr}
	   \eqno(34)$$
is asymptotically-flat, nonsingular outside the horizon and regulr
at the event horizon for
${\kappa \alpha^2 \over{M^2-a^2}} \ge 1 $.

\chapter{Discussion}
Some interesting characteristics for the structure of the axionic Kerr
black hole were studied in reference[5].
Besides the ring singularity at $r=0,~\theta
={\pi\over 2}$, there exist two extra singularities
$(\overline{r}_\pm =r\pm \mu\sqrt{M^2-a^2}~\,,~~~-1<\mu<1)$
between the inner event horizon $r_-$ and the
outer event horizon $r_+$. The extra singularities may be regarded as the
sources of axionic field strength. Therefore, if we are only interested
in the region outside the source,
the axionic Kerr black hole solution is an acceptable
physical solution. The non-vanishing axionic field strength could offer
observational
features which might not require a detail knowledge of the way that the
fundamental string theory devolves into the standard model.

It is worthwhile to note that the axisymmetric axion field strength
will still induce an axisymmetric non-rotating black hole if $a=0$,
and the axionic Kerr black hole will reduce to the Kerr black hole
if the new axionic hair is absent {\it i.e.} $\alpha=0$.  When we set both
$a=0$ and $\alpha =0$, the Schwarzschild black hole is reproduced
but with vanishing spherical symmetric Kalb-Ramond field.
However, it is, under a gauge transformation, consistent with those results
in reference [1,2,4] --- the unique spherical symmetric black hole
solution of EKR equation is Schwarzschild solution with non-vanishing
Kalb-Ramond field $B_{13}={q\over{4\pi r^2}}$.
In fact, the field strength  $H_{\mu\nu\lambda}$ is invariant under
the gauge transformation
$B^\prime_{\mu\nu}=
B_{\mu\nu}+\partial_\mu \Lambda_\nu-\partial_\nu \Lambda_\mu$.
Since
both of them give the same vanishing field strength $H_{\mu\nu\lambda}$,
we can recover the non-vanishing
Kalb-Ramond field $B_{13}={q\over{4\pi r^2}}$
by a gauge transformation $\Lambda_t=\Lambda_r=\Lambda_\theta=0$ and
$\Lambda_\phi=-{q\over 4\pi}\cos\theta$.

\vskip 2.5cm
\centerline{\bf Acknowledgements}
\vskip 0.5cm
We thank Prof. C.-R. Lee and Prof. H. T. Su for reading the manuscript.
This work was supported in part by the National Science Council of
the Republic of China under grant NCS-81-0208-M006-05.
\endpage

\def\jo{\journal}
\def\prl{Phys. Rev. Lett.}
\def\pr{Phys. Rev.}
\def\pl{Phys. Lett.}

\def\grg{Gen. Rel. Grav.}
\def\cqg{Class. Quantum Grav.}

\ref{M.J.~Bowick, S.B.~Giddings, J.A.~Harvey, G.T.~Horowitz and
     A.~Strominger \jo\prl&61(88)2823. }
\ref{M.J.~Bowick\jo\grg&22(90)137.}
\ref{B.A.~Campbell, M.J.~Duncan and K.A.~Olive\jo\pl&B251(90)34,{\bf B263}
    (1991) 364.}
\ref{R.R.~Hsu\jo\cqg&8(91) 779.}
\ref{R.R.~Hsu and W.F.~Lin\jo\cqg&8(91) L161.}
\ref{S.~Coleman, J.~Preskill and F.~Wilczek \jo{Mod. Phys. Lett.}&A6(91)
     1631.}
\ref{S.~Chandrasekhar,``{\it The Mathematical Theory of Black Hole}'',
  Oxford University Press (1983) Ch.6.}
\ref{see a similar argument used in R. Rauch and H. T. Nieh \jo\pr&D25
     (81) 2029, appendix B, and S. Weinberg, "{\it Gravitation and
     Cosmology}", New York, Wiley (1992) Ch. 13.}
\ref{D.C.~Robison\jo\prl&34(75)905.}

\refout
\end
\bye